\documentclass[cits]{PoS}

\usepackage{amsmath,bm}
\usepackage{tabularx}
\usepackage{slashed}

\newcommand{\Tr}{\mathrm{Tr}}
\newcommand{\STr}{\mathrm{STr}}
\newcommand{\I}{\mathrm{i}}
\newcommand{\Nf}{N_{\text{f}}}
\newcommand{\Nc}{N_{\text{c}}}

\newcommand{\separation}{\\[-2mm]}

\definecolor{blue}{rgb}{0,0,1}

\definecolor{green}{rgb}{0,1,0}

\definecolor{red}{rgb}{1,0,0}
\definecolor{RED}{rgb}{1,0,0}

\newcolumntype{L}[1]{>{\raggedright\arraybackslash}p{#1}} 
\newcolumntype{C}[1]{>{\centering\arraybackslash}p{#1}} 
\newcolumntype{R}[1]{>{\raggedleft\arraybackslash}p{#1}} 

\graphicspath{
{./}
{./figures/}
}

\newcommand{\TUD}{Theoriezentrum, Institut f\"ur Kernphysik, TU Darmstadt, 64289 Darmstadt, Germany}
\newcommand{\RKU}{Institut f\"ur theoretische Physik, Ruprecht-Karls-Universit\"at Heidelberg, 69120 Heidelberg, Germany}
\newcommand{\JLU}{Institut f\"ur theoretische Physik, Justus-Liebig-Universit\"at Gie\ss en, 35392 Gie\ss en, Germany}
\newcommand{\GSI}{GSI Helmholtzzentrum f\"ur 
  Schwerionenforschung GmbH, 64291 Darmstadt, Germany}
  
\title{Finite-Temperature Spectral Functions \\from the Functional Renormalization Group}

\ShortTitle{Finite-Temperature Spectral Functions from the FRG}

\author{\speaker{Ralf-Arno Tripolt}\\
        \TUD\\
        E-mail: \email{tripolt@theorie.ikp.physik.tu-darmstadt.de}
	     }

\author{Nils Strodthoff\\
        \RKU\\
        E-mail: \email{n.strodthoff@thphys.uni-heidelberg.de}}

\author{Lorenz von Smekal\\
        \TUD\\
        \JLU\\
        E-mail: \email{lorenz.smekal@physik.tu-darmstadt.de}}

\author{Jochen Wambach\\
        \TUD\\
        \GSI\\
        E-mail: \email{jochen.wambach@physik.tu-darmstadt.de}}

\abstract{
We present a method to obtain spectral functions at finite temperature 
from the Functional Renormalization Group. Our method is
based on a thermodynamically consistent truncation of the 
flow equations for 2-point functions with analytically continued
frequency components in the originally Euclidean external momenta. 
For the uniqueness of this continuation at finite temperature we 
furthermore implement the physical Baym-Mermin boundary conditions.
Results are presented for mesonic spectral functions obtained from a 
two-flavor quark-meson model.
}

\FullConference{31st International Symposium on Lattice Field Theory LATTICE 2013\\
                 July 29 - August 3, 2013\\
                 Mainz, Germany}

\begin{document}

\section{Introduction}

The computation of real-time observables represents a common challenge within lattice
calculations and other Euclidean approaches to Quantum Field Theory where an analytic 
continuation from imaginary to real time is needed for dynamic processes especially with
timelike momentum transfer. This technical difficulty arises already in the vacuum but is
even more severe at finite temperature where these continuations are based on discrete 
Matsubara frequencies and hence require additional boundary conditions for uniqueness
\cite{Baym1961,Landsman:1986uw}. Even with the analytic structure
completely fixed, however, the reconstruction of spectral functions
from discrete numerical (i.e., noisy) data on Euclidean
correlation functions, for example, is an ill-posed inverse
problem. Maximum entropy methods (MEM)
\cite{Jarrell:1996,Asakawa:2000tr,Karsch:2001uw,Datta:2003ww,Ding:2012sp}, Pad\'{e} approximants 
\cite{Vidberg:1977,Schmidt:2011,Dupuis:2009,Sinner:2009}, or very recently also a standard
Tikhonov regularization 
\cite{Dudal:2013yva} have been proposed to
deal with this problem. They all work best at low temperatures when
the density of Matsubara modes is sufficiently large, but they all
break down when the Euclidean input data is not sufficiently dense and
precise.  

Therefore any approach that can deal with the analytic 
continuation explicitly is highly desirable.
In this work we employ the Functional Renormalization Group (FRG) which has proven very useful in 
nonperturbative applications in quantum field theory and statistical physics, see e.g.\
\cite{Berges:2000ew,Polonyi:2001se,Pawlowski:2005xe,Schaefer:2006sr,Gies:2006wv,Kopietz2010,Braun:2011pp} 
for introductions. 
Within this framework the analytic continuation 
can be achieved already on the level of the flow equation, as proposed in 
\cite{Strodthoff:2011tz,Kamikado2013} and \cite{Floerchinger2012}.
In addition to its simplicity our approach enjoys a
number of particular advantages: First of all, it is thermodynamically
consistent in that the spacelike limit of zero external momentum in
the 2-point correlation functions agrees with the curvature or
screening masses as extracted from the thermodynamic grand potential
\cite{Strodthoff:2011tz}. Secondly, it satisfies the
physical Baym-Mermin boundary conditions  at finite temperature 
\cite{Baym1961} whose implementation here proceeds essentially as in a 
simple one-loop calculation \cite{Das:1997gg}. Finally, although we
focus on mesonic spectral functions obtained from the quark-meson model
\cite{Jungnickel:1995fp, Schaefer:2004en}
in this work it can be extended to
calculate also quark and gluonic spectral functions as an alternative
to analytically continued Dyson-Schwinger equations (DSEs)
\cite{Strauss:2012dg}, or to using MEM on Euclidean FRG
\cite{Haas:2013hpa}  or DSE results
\cite{Nickel:2006mm,Mueller:2010ah,Qin:2013ufa}.

\section{Flow Equations for the Quark-Meson Model}\label{sec:flow_equations}

\begin{figure}[t]
\centering
\includegraphics[width=0.5\columnwidth]{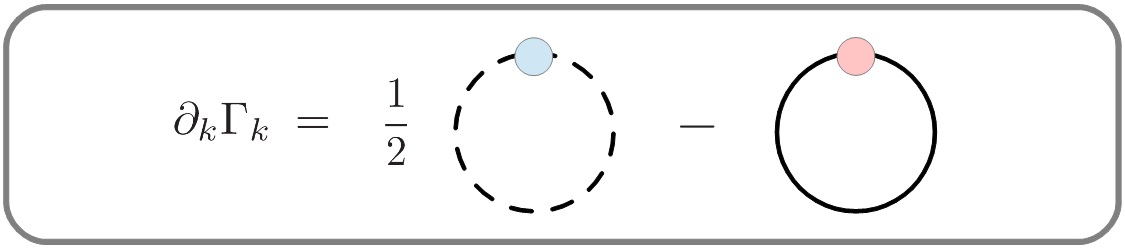}
\caption{(color online) Diagrammatic representation of 
the flow equation for the effective action. Dashed (solid) lines 
represent bosonic (fermionic) propagators and circles the respective regulator 
insertions $\partial_k R_k$.}
\label{fig:flow_Gamma} 
\end{figure}

Within the FRG approach pioneered by Wetterich \cite{Wetterich:1992yh} the effective action 
is generalized to the scale dependent effective average action $\Gamma_k$ via the introduction of
a regulator function $R_k$ which serves as an infrared (IR) regulator with an associated RG scale $k$.
This regulator has then to be removed by taking $k$ from the ultraviolet (UV)
cutoff scale $\Lambda$ down to zero. 
The scale dependence of the effective average action is governed by an exact one-loop equation
involving full scale- and field-dependent propagators which takes the form 
\begin{equation}
\label{eq:floweffaction}
\partial_k \Gamma_k=\tfrac{1}{2}\STr{[} \partial_k R_k(\Gamma^{(2)}_k+R_k)^{-1}{]},
\end{equation}
where $\Gamma^{(2)}_k$ denotes the second functional derivative of the
effective average action, and the supertrace includes internal and
spacetime indices, the functional trace, typically in  momentum space,
and the minus sign with degeneracy factor for fermion loops, see Fig.~\ref{fig:flow_Gamma}
for a diagrammatic representation.

In the following we employ the quark-meson model,
which serves as a low energy effective model for QCD with $\Nf=2$ light quark flavors 
and yields the following Ansatz for the effective average action, 
in the lowest order derivative expansion where only the effective potential carries
a scale  dependence,
\begin{equation} 
\Gamma_{k}[\bar\psi,\psi,\phi]= \int d^{4}x \Big\{
\bar{\psi} \left(\slashed{\partial} +
h(\sigma+\I\vec{\tau}\cdot\vec{\pi}\gamma_{5})\right)\psi
+\tfrac{1}{2} (\partial_{\mu}\vec\phi)^{2}+U_{k}(\phi^2) - c \sigma 
\Big\}\,,
\label{eq:QM}
\end{equation}
with $\phi_i=(\sigma,\vec \pi)_i$ and $\phi^2=\sigma^2+\vec \pi^2$. The effective potential
$U_{k}(\phi^2)$ allows for spontaneous breaking of chiral symmetry while the explicit breaking term
$c\sigma$ accounts for a non-vanishing pion mass.

\begin{figure}[h]
\centering
\includegraphics[width=0.49\columnwidth]{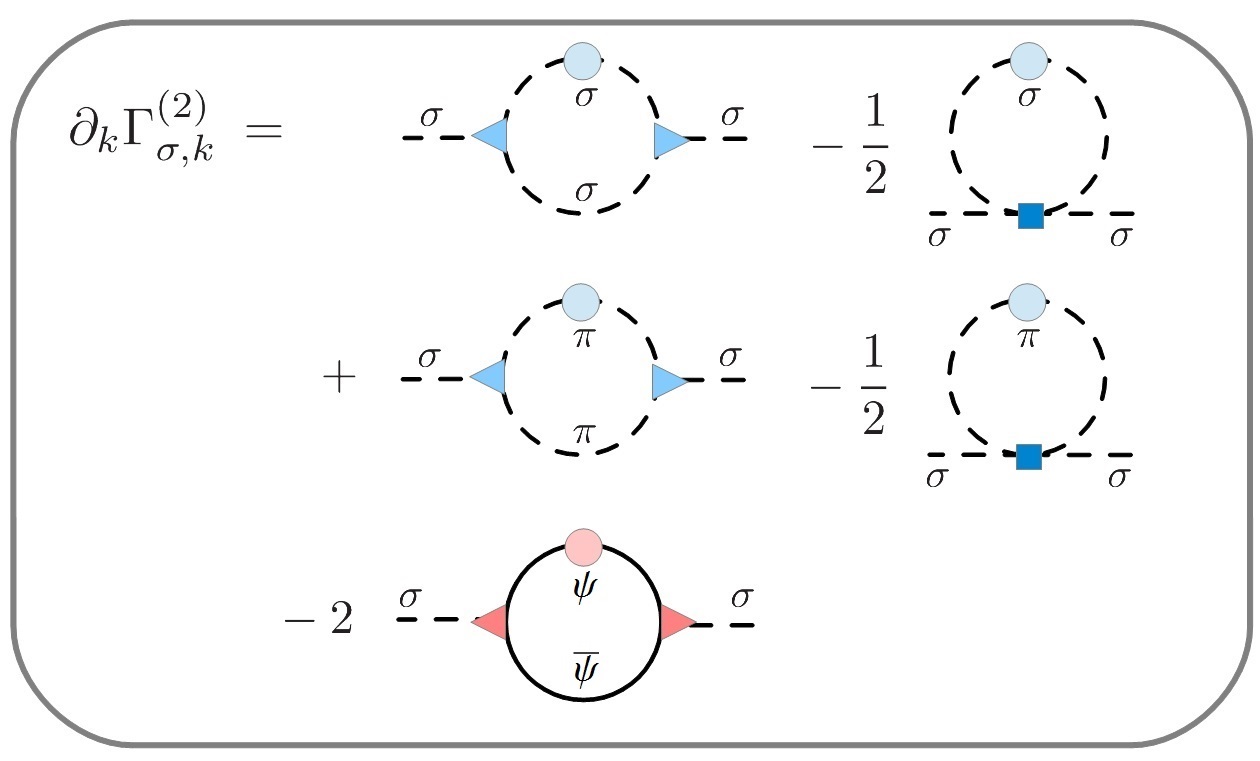}
\includegraphics[width=0.50\columnwidth]{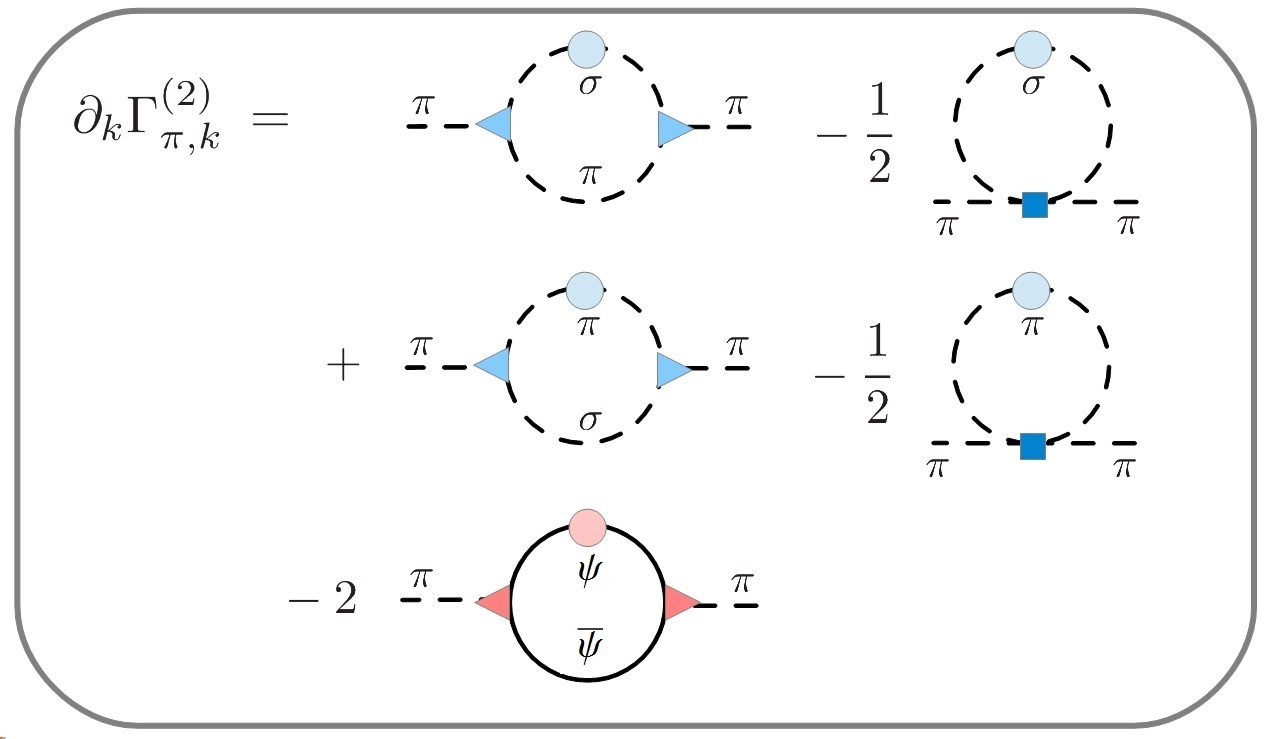}
\caption{(color online) Diagrammatic representation of 
the flow equations for the inverse mesonic 2-point functions.
Three-point vertices are represented by triangles,
four-point vertices by squares.}
\label{fig:flow_Gamma2} 
\end{figure}

Inserting Eq. (\ref{eq:QM}) into the flow equation for the effective action,
Eq. (\ref{eq:floweffaction}), evaluated for constant fields and using the three-dimensional 
analogues of the LPA-optimized regulator functions \cite{Litim:2001up},
gives for the flow equation of the effective potential
\begin{equation}
\label{eq:flow_pot} 
\partial_k U_k =
\tfrac{1}{2} I_{\sigma}^{(1)} +
\tfrac{1}{2}(N-1) I_{\pi}^{(1)} -
\Nc \Nf I_{\psi}^{(1)}
.
\end{equation}
Therein the loop functions $I_{\alpha}^{(i)}$ are defined as
\begin{align}
\label{eq:I_def} 
I_{\alpha}^{(i)} = \Tr_q 
\left[ 
\partial_k R_k(q)
G_{\alpha,k}(q)^{i}
\right],
\end{align}
with $\alpha\in\{\sigma,\pi,\psi\}$, $G_{\alpha,k}(q)$ the full (scale-dependent) Euclidean propagator
and $R_k(q)$ chosen appropriately for bosonic and fermionic fields.
The flow equations for the inverse mesonic 2-point functions are now
obtained by taking two functional derivatives of Eq.~(\ref{eq:floweffaction}),
\begin{align}
\label{eq:gamma2sigma}
\partial_k \Gamma^{(2)}_{\sigma,k}&=
J^B_{\sigma\sigma}(\Gamma_{\sigma\sigma\sigma}^{(0,3)})^2
+(N-1) J^B_{\pi\pi} (\Gamma_{\sigma\pi\pi}^{(0,3)})^2 
-\tfrac{1}{2}I_\sigma^{(2)}\Gamma_{\sigma\sigma\sigma\sigma}^{(0,4)}
-\tfrac{(N-1)}{2} I_\pi^{(2)}\Gamma_{\sigma\sigma\pi\pi}^{(0,4)}
-2\Nc \Nf J^F_{\sigma}
,
\\
\label{eq:gamma2pion}
\partial_k \Gamma^{(2)}_{\pi,k}&=
(J^B_{\sigma\pi} +J^B_{\pi\sigma})(\Gamma_{\sigma\pi\pi}^{(0,3)})^2-\tfrac{1}{2}
I_\sigma^{(2)}\Gamma_{\sigma\sigma\pi\pi}^{(0,4)}
-\tfrac{1}{2}I_\pi^{(2)}\!(\Gamma_{\pi\pi\pi\pi}^{(0,4)}
+\!(N\!-\!2)\Gamma_{\pi\pi\tilde{\pi}\tilde{\pi}}^{(0,4)})
-2\Nc \Nf J^F_{\pi}
,
\end{align}
with $\pi\neq\tilde{\pi}\in \{\pi_1,\pi_2,\pi_3\}$ and $N=4$ in our $O(4)$ case, see 
Fig.~\ref{fig:flow_Gamma2} for a graphical representation. 
The bosonic and fermionic loop functions are defined as
\begin{align}
\label{eq:J_def}
J^B_{\alpha\beta}(p)=\:&\Tr_q
\left[ 
\partial_k R_{k}(q)G_{\alpha,k}(q-p)G_{\beta,k}(q)^2
\right], \\
\label{eq:J_F_def}
J^F_{\alpha}(p)=\:&\Tr_q 
\left[ 
\partial_k R_{k}(q)G_{\psi,k}(q)
\Gamma_{\bar{\psi}\psi\alpha}^{(2,1)}
G_{\psi,k}(q-p)\Gamma_{\bar{\psi}\psi\alpha}^{(2,1)}G_{\psi,k}(q)
\right],
\end{align}
for $\alpha\in\{\sigma,\pi\}$ and at external momentum $p$. All vertices are taken
to be momentum independent in our present truncation and are obtained by appropriate
functional derivatives of the effective action for the quark-meson model, given by Eq. (\ref{eq:QM}). 

\vspace{5mm}

\section{Analytical Continuation and Numerical Implementation}
\label{sec:continuation}

We employ the following two-step procedure for the
analytic continuation from imaginary to real time:  
Once the sum over the Matsubara frequencies in the flow equations for
the 2-point functions is performed,  we first exploit the periodicity of the
bosonic and fermionic occupation numbers along the imaginary direction of
the complex energy plane, i.e.\ with respect to discrete
Euclidean external Matsubara modes $p_0=2n\pi T$,
\begin{equation}
n_{B,F}(E+\I p_0)\rightarrow n_{B,F}(E), 
\end{equation}
cf. \cite{Tripolt2013a} for explicit expressions of the flow equations. 
In the second step, the retarded
2-point functions are then obtained from their Euclidean counterparts
via the analytic continuation
\begin{equation}
\Gamma^{(2),R}(\omega,\vec p)=\lim_{\epsilon\to 0} \Gamma^{(2),E}
(p_0=-\I(\omega+\I\epsilon), \vec p),
\end{equation}
where we
keep a small but finite value of $\epsilon = 1\,{\rm MeV}$ in our numerical 
implementation and consider the case of vanishing spatial momentum, $\vec p = 0$.
This substitution of the discrete Euclidean external $p_0$ by the
continuous real frequency $\omega $ is done explicitly within the flow equation, 
before the 
integration of the RG scale $k$. Because of the one-loop structure of
the flow equations together with the unregulated Matsubara sums, the
correctness of this analytic continuation in the complex frequency
plane follows directly from the corresponding one-loop formulae for
the polarization functions in thermal field theory \cite{Das:1997gg}.

Finally, the spectral function is given by the discontinuity of the
propagator and can hence be expressed in terms of  
the imaginary part of the retarded propagator as,
\begin{equation}
\rho(\omega)=-\frac{1}{\pi}\frac{\text{Im}\,\Gamma^{(2),R}(\omega)}{\left(\text{Re}\,
\Gamma^{(2),R}(\omega)\right)^2+\left(\text{Im}\,\Gamma^{(2),R}(\omega)\right)^2}.
\end{equation}

In order to numerically solve the flow equations for the effective potential and for
the real and imaginary parts of the 2-point functions we take the effective potential
at the UV scale to be of the form
\begin{equation}
\label{eq:initial_conditions} 
U_\Lambda(\phi^{2}) =
\tfrac{1}{2}m_\Lambda^{2}\phi^{2} +
\tfrac{1}{4}\lambda_\Lambda(\phi^{2})^{2}
\,,
\end{equation}
where the chosen values for the parameters and the corresponding values for observables in the IR
are summarized in Table \ref{tab:parameters}. For further details on our numerical implementation
see \cite{Tripolt2013a}.

\begin{table}[t]
\centering
\begin{tabular}{C{1.2cm}C{1.2cm}C{1.1cm}C{1.2cm}C{1.0cm}|C{1.3cm}C{1.3cm}C{1.3cm}C{1.3cm}}
 $\Lambda$/MeV & $m_\Lambda/\Lambda$ & $\lambda_\Lambda$ & $c/\Lambda^3$ &  $h$  
 & $f_\pi$/MeV  & $m_\pi$/MeV  & $m_\sigma$/MeV & $m_\psi$/MeV \\
\hline
1000 & 0.794 & 2.00 & 0.00175 & 3.2 & 93.5 
 & 137 & 509 & 299
\end{tabular}
\caption{Employed parameter set and obtained vacuum values for observables in the IR, 
i.e. $k_{{\rm IR}}\approx 20 \,{\rm MeV}$.}
\label{tab:parameters} 
\end{table}

\section{Results}

\begin{figure}[b]
\centering
\qquad\includegraphics[width=0.6\columnwidth]{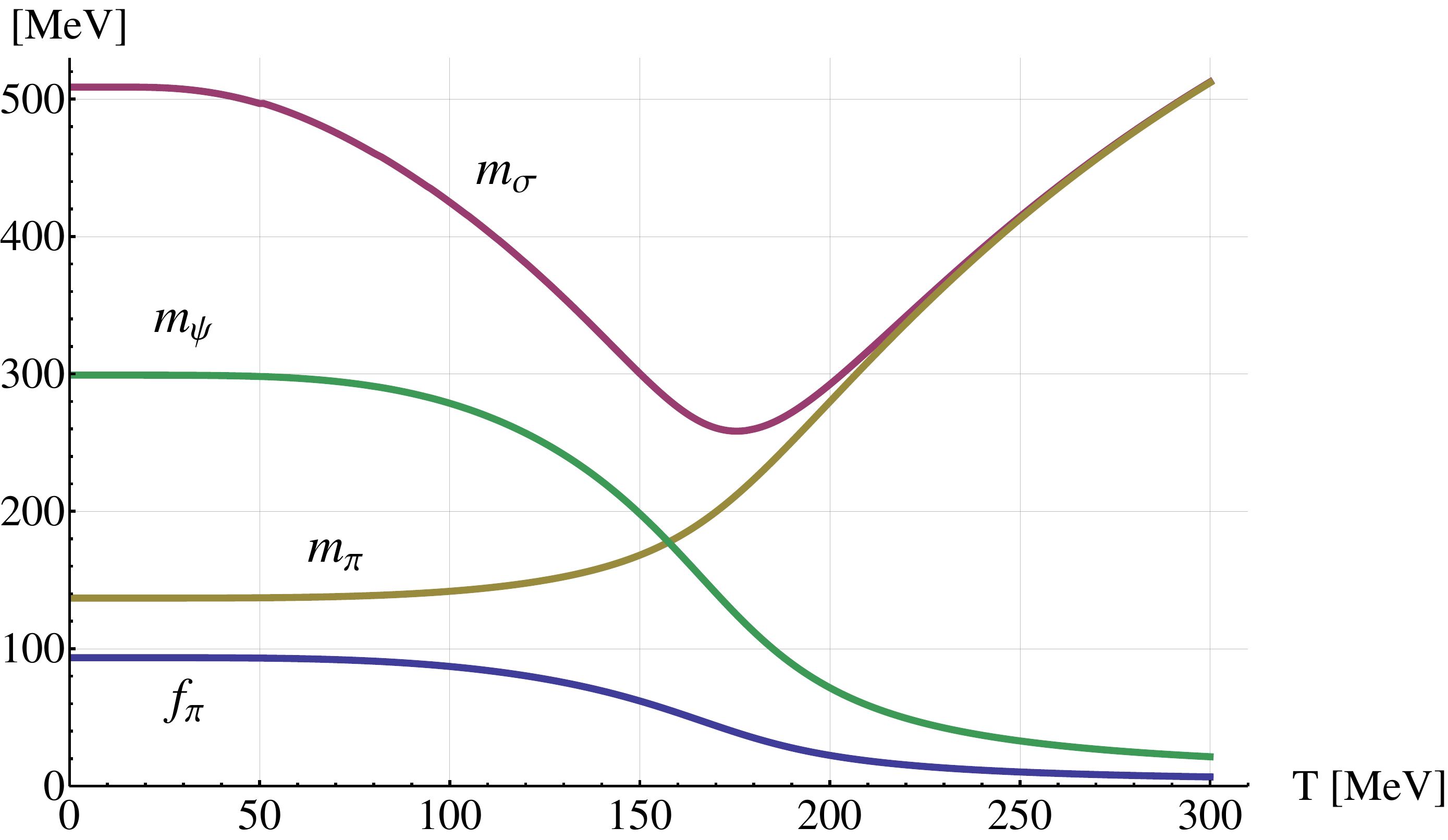}
\caption{(color online) Meson masses, quark mass and chiral order parameter vs. temperature.}
\label{fig:masses} 
\end{figure}

Since the pion and sigma meson spectral functions are strongly affected by the possible
decay processes, which in turn depend on the masses of the involved particles, 
we first turn to a discussion of the temperature-dependence of the meson screening masses and the quark mass, 
cf. Fig.~\ref{fig:masses}. One clearly observes the expected crossover behavior with a 
pseudo-critical temperature of $T_c\approx 175\, {\rm MeV}$ as obtained from the maximum of the chiral 
susceptibility $\chi_\sigma\equiv 1/m_\sigma^2$, see 
\cite{Schaefer:2006ds,Tripolt2013} and \cite{Strodthoff:2013cua} for a 
comparison of different definitions of pseudo-critical temperatures. At 
higher temperatures the sigma and pion mass become degenerate, while the quark 
mass and the order parameter drastically decrease, indicating the progressing restoration of chiral symmetry. 

\begin{figure}[t]
\includegraphics[width=0.49\linewidth]{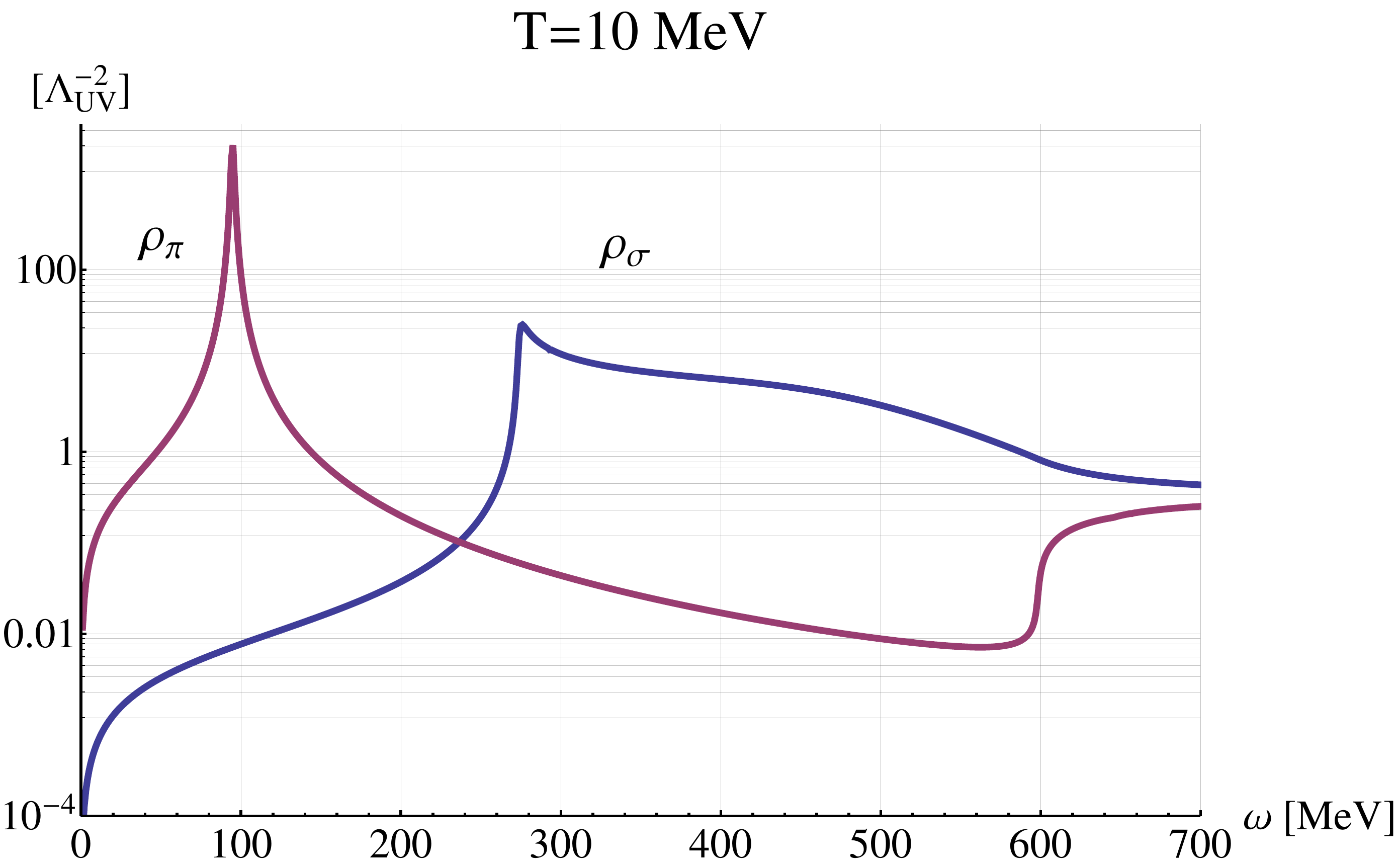}
\includegraphics[width=0.49\linewidth]{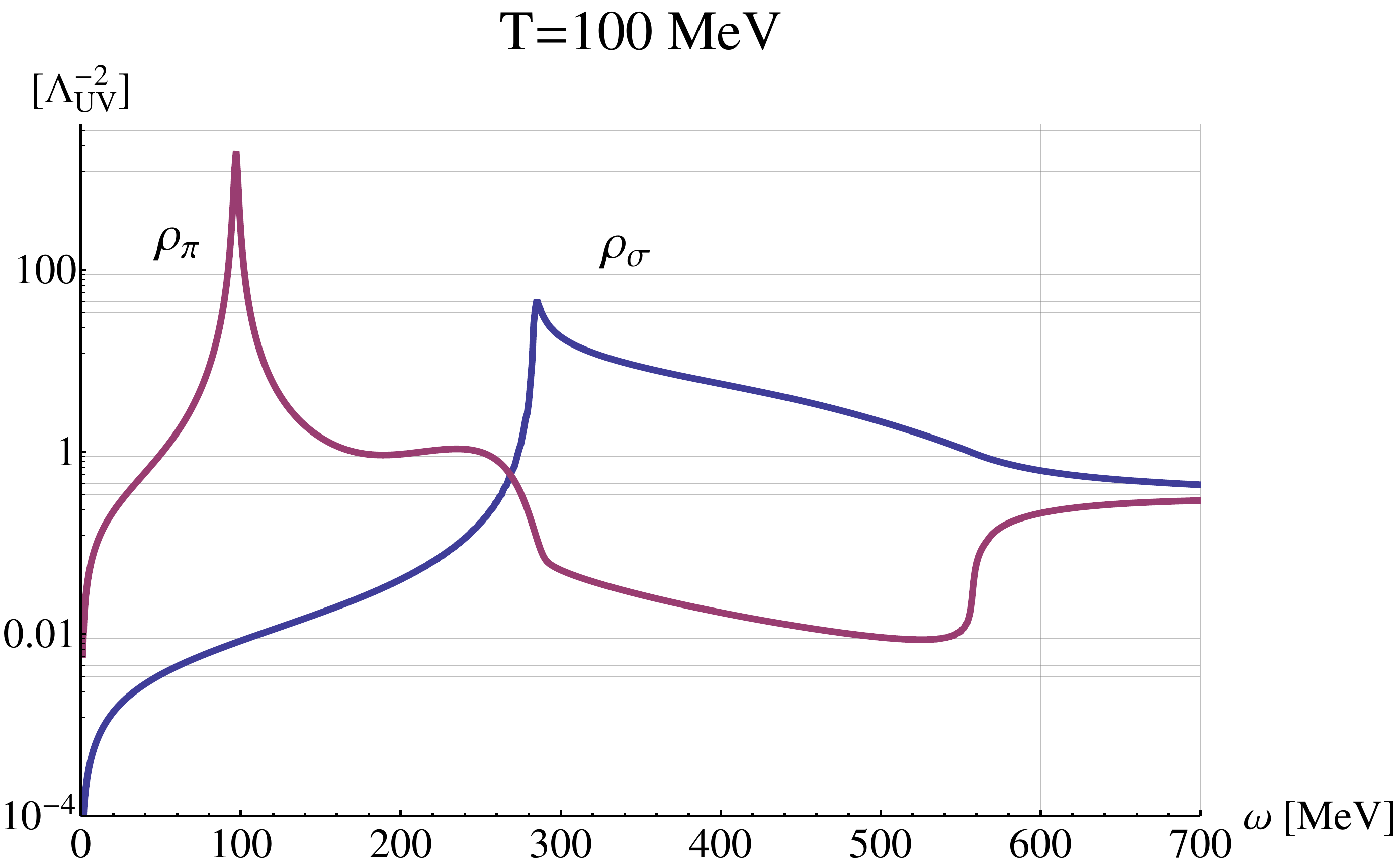}\\[2mm]
\includegraphics[width=0.49\linewidth]{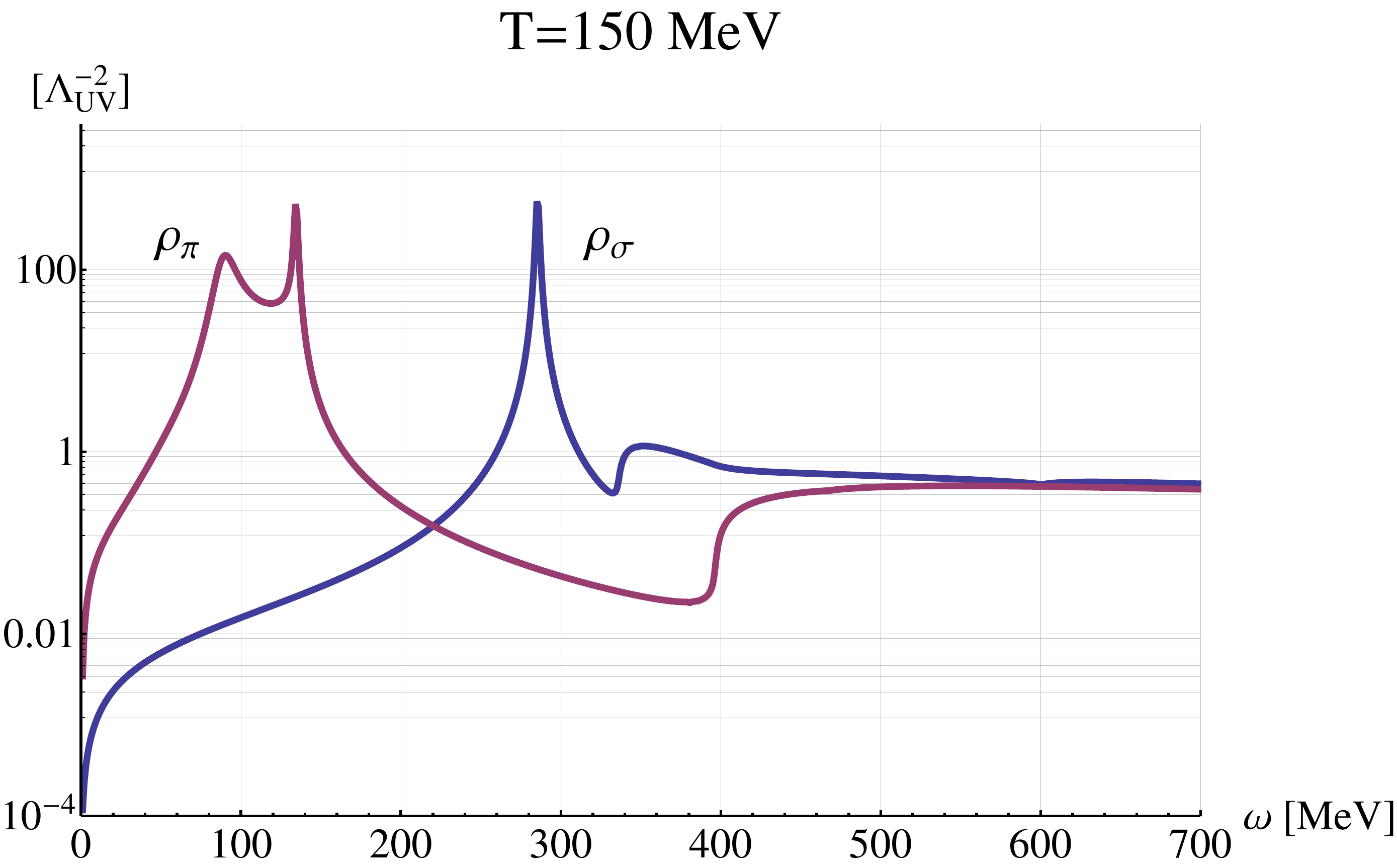}
\includegraphics[width=0.49\linewidth]{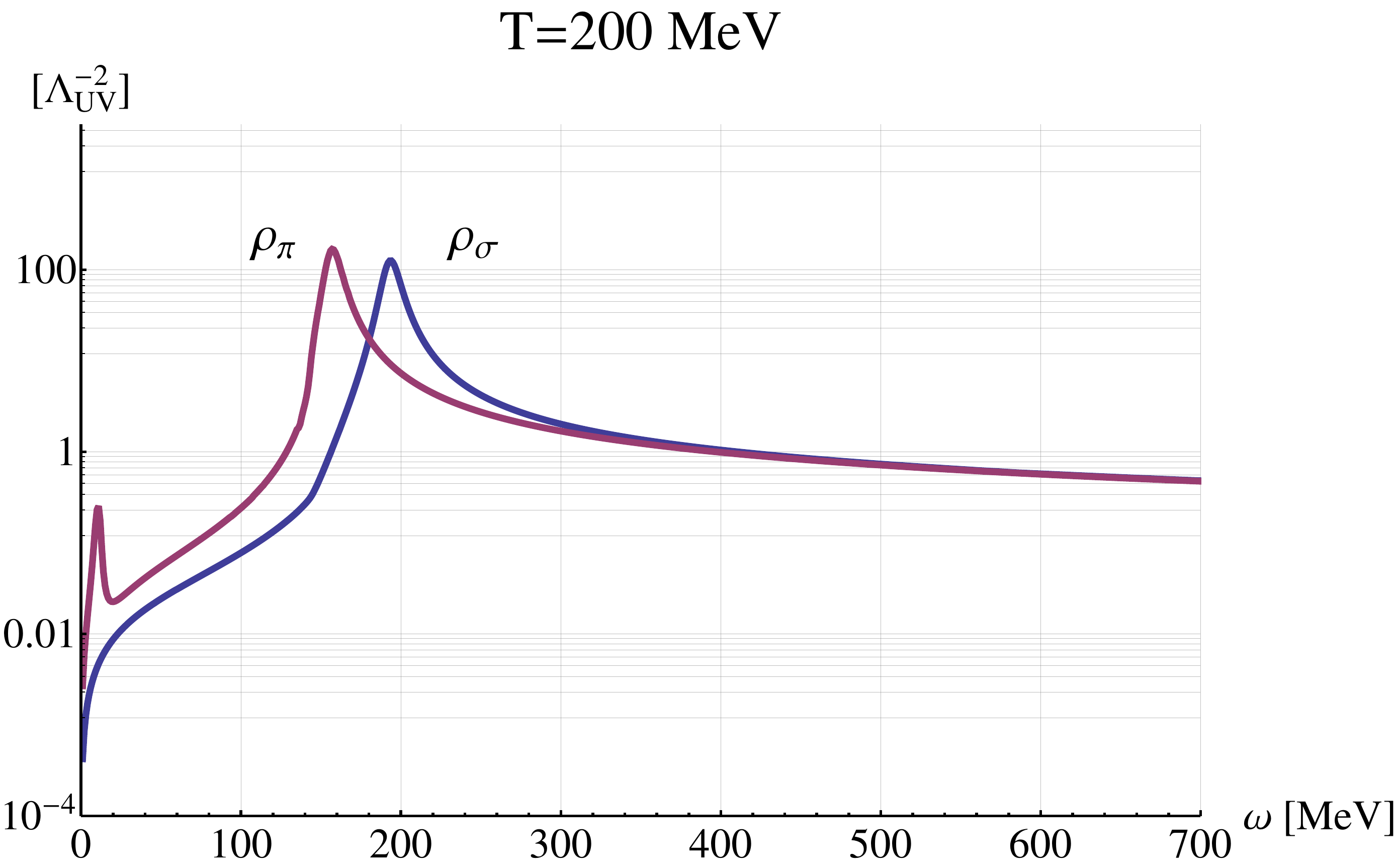}
\caption{(color online) Sigma and pion spectral function vs. external energy at different temperatures. 
}
\label{fig:spectralfunctions} 
\end{figure}

We now turn to a discussion of the pion and sigma meson spectral function
which are shown as a function of external energy at different temperatures in
Fig.~\ref{fig:spectralfunctions}.
At $T=10\,{\rm MeV}$ the spectral functions closely 
resemble the vacuum structure already observed in previous studies, 
cf.\ \cite{Kamikado2013,Kamikado2013a}. 
For external energies larger than $2m_\psi$ the decay of an (off-shell) pion into a quark anti-quark pair,
${\pi'\rightarrow \bar{\psi} \psi}$, becomes 
energetically possible, giving rise to an increase of the pion spectral function at
$\omega\geq 2m_\psi \approx 600\,{\rm MeV}$.
The dominant channel affecting the sigma spectral function is the decay into two pions, 
${\sigma'\rightarrow \pi\pi}$, which is possible for $\omega\geq 2m_\pi$ 
and renders the sigma meson unstable. 

At higher temperatures the process ${\pi'\pi\rightarrow \sigma}$, which describes an off-shell pion and a pion of the heat bath
going into a sigma meson, contributes to the pion spectral function for 
$\omega\leq m_\sigma-m_\pi$. At $T=150\,{\rm MeV}$ this results in modifications of $\rho_\pi$
near $\omega\approx 100\,{\rm MeV}$ while the sigma spectral function develops a sharp peak at
$\omega\approx 280\,{\rm MeV}$, as expected near the chiral crossover, where neither the decay 
into two pions nor into two quarks is energetically possible. 
When increasing the temperature further, 
the quarks become the lightest degrees of freedom considered here, providing decay channels 
for both the pion and the sigma meson down to small external energies, leading to a 
broadening of the pronounced peaks in the spectral functions which eventually become degenerate.

\section{Summary and Outlook}
We have presented a method to obtain spectral functions at finite temperature 
from the Functional Renormalization Group, based on previous 
studies in the vacuum \cite{Kamikado2013,Kamikado2013a}. 
The method involves an analytic continuation
from imaginary time to real time performed on the level of the flow equations and has been 
illustrated here for mesonic spectral functions obtained from the 
quark-meson model. 
Our approach is comparably simple and can be 
extended in several ways, for example by including finite quark chemical potential 
which will allow to study spectral functions over the whole range of the phase diagram, 
including critical regimes, cf. \cite{Tripolt2013a}.
Additionally, an extension to non-vanishing external spatial 
momenta is possible and will allow for the calculation of transport coefficients such as the shear viscosity. 

\acknowledgments 
The authors thank Kazuhiko Kamikado and Jan Pawlowski for 
discussions and work on related subjects. This work was supported by
the Helmholtz International Center for FAIR within the LOEWE
initiative of the State of Hesse. L.v.S.~is furthermore supported 
by the European Commission, FP-7-PEOPLE-2009-RG, No. 249203, N.S.
by the grant ERC-AdG-290623, and \mbox{R.-A.T.} by the 
Helmholtz Research School for Quark Matter Studies, H-QM.

\clearpage

\end{document}